\documentstyle[12pt]{article}

\begin{document}
\setcounter{footnote}{0}
\setcounter{equation}{0}
\setcounter{figure}{0}
\setcounter{table}{0}
\vspace*{5mm}

\begin{center}
{\large\bf Diffusion Poles and the Anderson Transition }

\vspace{4mm}
I. M. Suslov \\
P.L.Kapitza Institute for Physical Problems,
\\
119337 Moscow, Russia \\
\vspace{4mm}

\begin{minipage}{135mm}
{\rm In the recent series of papers (cond-mat/0402471,
cond-mat/0403618, cond-mat/0407618, cond-mat/0501586),
Janis and Kolorenc discussed the role of the diffision
poles in the Anderson transition theory. Their picture
contradicts the general principles and is shown below to
be completely misleading.  Correct setting of the problem
is given and the contemporary situation is discussed.
The critical remarks are given on the relation of the
diffusion coefficient with multifractality of the wave functions. }
\end{minipage} \end{center}

\vspace{6mm}

The present paper has an aim to discuss the controversy in the
literature concerning the role of the diffision poles in the Anderson
transition theory.
\vspace{3mm}

1.  The spatial dispersion of the diffusion coefficient $D(\omega, q)$
has a fundamental significance for the whole Anderson transition theory.

It is well known (see e.g. \cite{1,2}) that a quantity
$$
\phi^{RA} ({\bf r- r'}) = \langle G_{E + \omega}^R
({\bf r}, {\bf r'}) G_E^A ({\bf r'} {\bf r}) \rangle
\eqno(1)
$$
in the momentum representation has a diffusion pole
$$
\phi^{RA} ({q}) = \frac{2 \pi \nu (E)}{-i \omega + D(\omega, {q}) q^2} +
\phi_{reg} ({q}),\eqno(2)
$$
where  $D(\omega, q)$ is the observable diffusion coefficient, $G^R$
and $G^A$ are nonaveraged retarded and advanced Green functions for
an electron in the random potential, $\nu(E)$ is a density of states
for the
energy $E$. In the localized phase, the diffusion pole transforms into
the Berezinskii-Gorkov singularity
$$
\phi^{RA} ({q}) = \frac{2 \pi \nu(E)}{- i \omega} A ({q}) + \phi_{reg}
({q}),\eqno(3)
$$
$$
A ({q}) = \int d {\bf r} e^{-i {q} {\bf r}} A ({\bf r})\,,
\qquad
A ({\bf r}) = \frac{1}{\nu(E)} \left\langle \sum \limits_s
| \psi_s ({\bf r})|^2 | \psi_s (0)|^2 \delta (E - \epsilon_s)
\right\rangle ,\eqno(4)
$$
where $\psi_s({\bf r})$ and $\epsilon_s$ are eigenfunctions and
eigenenergies of the electron in  the random potential.
In the original paper \cite{3},  the  $\delta (\omega)$ singularity
(occuring from the terms with $s=s'$)
was established for the density correlator
$$
S({\bf r-r'})=\frac{1}{\nu(E)} \left\langle  \sum\limits_{s,s'}
\psi_s^*({\bf r}) \psi_{s'}({\bf r'}) \psi_{s'}^*({\bf r'})
\psi_s({\bf r}) \delta (E - \epsilon_s)
\delta (E+\omega - \epsilon_{s'}) \right\rangle \eqno(5)
$$
but it
can be easily transform into the $1/\omega$ singularity for
$\phi^{RA}$ \cite{4,5,2}.  Comparison of (2) and (3) shows
that $D(\omega,{q}) \sim \omega$ in the localized phase: a slower
dependence would destroy the $1/\omega$ singularity in Eq.\,3 and
a more rapid dependence would lead to disappearance
of the  $q$-dependence  in the singular part of (2),
which surely exists according to Eqs.\,3,\,4. As a result,
$$
D (\omega, {q}) = - i \omega \,d (q)\,,
\eqno (6)
$$
where the limit $\omega \to 0$ is taken in the function $d(q)$.
It is easy to see from the relation
$$
\frac{1}{1 + d(q)q^2} = A ({q}) = \int d {\bf r} e^{-i {q}
\cdot {\bf r}} A ({\bf r}) \eqno(7)
$$
that $d(q)$ is a regular function of $q^2$, since all coefficients
in the expansion of (7) in the powers of $q^2$ are finite
due to exponential decay of $A ({\bf r})$ at large $r$. The analogous
regularity in $q^2$ is expected for $D (0, {q})$ in the metallic
state, while the anomalous spatial dispersion of the type $q^\alpha$
can occur at the transition point.

One can see from Eq.\,6, that $D (0, {q}) \equiv 0$ in the localized
phase; so the Anderson transition does not reduce to vanishing of
$D(0,0)$ but has essentially more deep nature \cite{8}. The question
arises, how the spatial dispersion of $D$ changes near the Anderson
transition. One can suggest, that $D(0,{q})$ vanishes at
the transition point simultaneously for all ${q}$. Such possibility
looks incredible from viewpoint of phenomenological considerations
in the spirit of the Landau theory. Indeed, all expansion
coefficients of $D(0,q)$ in $q^2$ should vanish simultaneously,
irrespective of the way the critical point is approached and its
location on the critical surface.  Obviously,
such vanishing cannot occur by chance,
and should be backed by some profound symmetry. Does
such a symmetry exist, and what is its origin?  If this
symmetry is taken for granted, then the order parameter for the Anderson
transition should have an infinite number of components.
Another possibility suggests, that $D(0, q)$ vanishes for a
single value of $q$, and then instability develops (in the spirit
of a soft mode) leading to a first-order phase transition. In this case,
one has to suggest an appropriate scenario. One can see, that even
a type of the Anderson transition cannot be understood without
solving the problem of the spatial dispersion of the diffusion
coefficient.

The situation is aggravated to the utmost by the existence of
the Ward identity \cite{1}
$$
\Delta \Sigma_{\bf k} ({\bf q}) = \frac{1}{N} \sum \limits_{{\bf k}_1}
U_{{\bf kk}_1} ({\bf q}) \Delta G_{{\bf k}_1} ({\bf q}),\eqno(8)
$$
$$
\Delta G_{\bf k} ({\bf q}) \equiv G_{{\bf k} + {\bf q}/2}^R - G_{{\bf k} -
{\bf q}/2}^A\,,  \qquad
\Delta \Sigma_{\bf k} ({\bf q}) \equiv \Sigma_{{\bf k} + {\bf q}/2}^R -
\Sigma_{{\bf k} - {\bf q}/2}^A\eqno(9)
$$
where $\Sigma_{\bf k}$ is the self energy and $U_{{\bf kk'}} ({\bf q}) $ is
the irreducible 4-vertex appearing in the Bethe--Salpeter equation.
This vertex has the diffusion pole for
${\bf k}+{\bf k}'\to 0$ \cite{1,2}
$$
U_{{\bf k} {\bf k}^\prime} ({\bf q}) =  U_{{\bf k} {\bf
k}^\prime}^{reg} ({\bf q}) +
 \frac{F ({\bf k}, {\bf k}^\prime, {\bf
q})}{-i \omega + D (\omega, {\bf k} + {\bf k}^\prime)({\bf k} + {\bf
k}^\prime)^2}\eqno(10)
$$
due to time-reversal invariance.
The left-hand side of Eq.\,8 is regular at the transition point,
whereas the integrand on the right-hand side diverges as
$1/\omega$ in the localized phase for all ${\bf k}, {\bf k}'$ in the
$\omega\to 0$ limit. This singularity should be cancelled after
integration over $ {\bf k}'$, which involves $D(\omega,{\bf k}+{\bf k}')$
and imposes stringent requirements on the approximation used for
calculation of the spatial dispersion of the diffusion coefficient.

\vspace{3mm}

2. The first attempt to deal with these problems was made
by Vollhardt and W${\rm {\ddot o}}$lfle \cite{1}.
They used approximation for $U_{{\bf kk}^\prime} ({\bf q})$
suggested by the weak localization theory \cite{9}
(corresponding to
$U_{{\bf kk}^\prime}^{reg} ({\bf q}) = \mbox{const}$,
$ F ({\bf k}, {\bf k}^\prime, {\bf q}) = \mbox{const}$ in Eq.\,10)
and solved approximately the Bethe--Salpeter equation.
Their solution corresponds to the following simple estimate.
The quantity $U_{{\bf kk}^\prime} ({\bf q})$ plays
the role of a 'transition probability' in the quantum kinetic equation,
and one can use the analogue of the $\tau$-approximation,
$D \propto l \propto \langle U \rangle ^{-1}$ ($l$ is the mean free path,
$\langle ...  \rangle$ denotes averaging over the momenta),
to obtain the relation
$$
D \sim {\rm const} \left ( U_0 + F_0 \int \frac{d^dq}{-i \omega +
D(\omega, q)q^2} \right )^{-1} \,,\eqno(11)
$$
coinciding with the self-consistency equation of the
Vollhardt and W${\rm {\ddot o}}$lfle theory.
With the increase of disorder, the 'transition probability' has anomalous
growth due to diminishing of the diffusion coefficient and provides
the possibility for vanishing of the latter.
If the spatial dispersion of $D (\omega, {q})$ is neglected, Eq.\,11
becomes closed and allows to determine the critical
exponents for the conductivity $\sigma$ and the
localization length $\xi$,
$$
\sigma \sim \tau^s, \qquad \, \, \, \xi \sim \tau^{- \nu}\eqno(12)
$$
($\tau$ is the distance to the transition). Setting
$D = \mbox{const} (\omega) \sim \sigma$ in the metallic phase
and $D \sim (-i \omega) \xi^2$ in
the localized phase, one can obtain
$$
s = 1\,\,, \quad d > 2\,\,;
 \qquad \qquad  \nu = \left \{ \begin{array}{cc}
{\displaystyle 1/(d-2)}\,\,,  & 2 < d < 4 \\ {   } & {  } \\
{\displaystyle 1/2} \,\,, & d > 4
\end{array} \right . \quad.
\eqno(13)
$$
Thus, the Vollhardt and W${\rm {\ddot o}}$lfle theory \cite{1} provides
the adequate qualitative description of the Anderson transition.
However, due to neglecting the spatial dispersion of $D (\omega, {q})$
the problem of the order parameter symmetry  was overlooked, while
approximations
$U_{{\bf kk}^\prime}^{reg} ({\bf q}) = \mbox{const}$,
$ F ({\bf k}, {\bf k}^\prime, {\bf q}) = \mbox{const}$
lead to the crude violation of the Ward identity.

The problem of the spatial dispersion of the diffusion coefficient
manifests itself also in the $\sigma$-models formalism \cite{10},
which is usually considered as the most rigorous approach to
the localization theory. These models are
derived by using the saddle-point approximation for integration
over 'hard' degrees of freedom in the functional integral and
subsequent expansion in gradients.  The 'minimal' $\sigma$-model
contains only the lowest (second) powers of gradients and corresponds
to neglecting the spatial dispersion of the diffusion coefficient.
One can suspect that it violates self-consistency of the theory.
Attempt to take the spatial dispersion of $D(\omega,{q})$ into
account and include the terms with higher gradients into the Lagrangian
of the $\sigma$-model leads to anomalous growth of these terms in
the course of the renormalization-group transformations \cite{11}:
the so called `high-gradient catastrophe' occurs. The stable fixed
point, corresponding to the true spatial dispersion, is still
not found.

The problem of the Ward identity was completely solved in the paper
\cite{2}. The crucial point consists in the use of the spectral representation
for the quantum collision operator $\hat L$ (which is the quantum analogue of
the Boltzman collision operator). Then cancellation of the $1/\omega$
singularity in the right-hand side of Eq.\,8 follows from the orthogonality
relations for eigenvectors of $\hat L$. At present, the spectral representation
for the operator $\hat L$ looks as a necessary ingredient of any consistent
theory of the Anderson transition: it is the only place where approximations
can be safely made without violation of the Ward identity.

The paper \cite{2} deals also with the general problem of evaluation
$D(\omega,{q})$ near the Anderson transition. It appears, that the
spectrum of the operator $\hat L$ possesses the non-trivial hierarhial
structure. The condition of stability of this hierarhy in respect to
infinitismal perturbations can be expressed in a form of a self-consistency
equation, which replaces the crude Eq.\,11. Solution of this equation is seeked
for an arbitrary form of the spatial dispersion of $D(\omega,{q})$, but
self-consistency is reached only for the solution with the weak $q$-dependence.
As a consequence, estimation of the integral in Eq.\,11 remains unchanged and
and we return to the result (13) for the critical exponents.

The paper \cite{2} takes into account only evident symmetry of the
system and considers a situation of the general position, compatible
with this symmetry. Such reasoning is typical for the mean field
theory and may be wrong due to existence of hidden symmetry.
However, there are arguments that the theory \cite{2} is
something more than the mean field theory. Indeed, the existence of
hidden symmetry is expected only for the critical point itself;
as a result, the mean-field theory does not give the true critical
behavior, but correctly describes the change of symmetry. From
viewpoint of the change of symmetry, the Anderson transition
is similar to the Curie point for an isotropic
$n$-component ferromagnet in the limit of $n\to\infty$:  it is
related with simultaneous vanishing of all coefficients in
the expansion of $D(0,q)$  over $q^2$.  The model of an
infinite-component ferromagnet is exactly solvable \cite{12}
and its critical exponents appear to be in complete agreement
with the results (13) of the straightforward analysis.
This is an argument, that the symmetry of the critical point is
established correctly and that the critical exponents are
determined exactly.

Another argument is given by the fact that Eq.\,13 agrees
with practically all reliable analytical results obtained for
specific models \cite{2,13}, and with experimental results
$s\approx 1$, $\nu\approx 1$ for $d=3$ \cite{34,35} obtained
independently in the measurements of conductivity and
the dielectric constant. It agrees also with the one-parameter
scaling hypothesis \cite{24}, though not one word about scaling
was said in derivation of the self-consistency equation.
Unfortunately, the results (13) are in a serious contradiction
with numerical experiments \cite{14}, though the latter have
their own problems \cite{15}, being in conflict not only with
(13) but also with some rigorous theorems.

In conclusion, a scenario of the Anderson transition suggested
in Ref.\,2 cannot be considered as rigorously established but has
the good perspectives to be exact. At present, it is
the only scheme which  makes it possible to analyse the spatial
dispersion of the diffusion coefficient near the critical point.

\vspace{3mm}

3. In the current literature, behavior of the diffusion
coefficient near the Anderson transition is discussed in
terms of one-parameter scaling \cite{22}, and
the following ansatz is used for a critical point \cite{23}
$$
D(\omega, q) \sim \omega^{\eta/d} q^{d-2-\eta}\,.
\eqno(14)
$$
The exponents in Eq.\,14 are chosen in such way that the
static diffusion coefficient $D_L$ of a finite system of size
$L$, determined from $D(\omega, q)$ by relation
$$
D_L \sim D(D_L/L^2, 1/L)\,,
$$
has a behavior $D_L\sim L^{2-d}$
predicted by the one-parameter scaling theory \cite{24}.
Relation $\eta=d-D_2$ was claimed \cite{25}, connecting
the exponent $\eta$ with the fractal dimension $D_2$ of wave
functions. Numerical results for $\eta$ and $D_2$ in three
dimensions have a large scattering ( $D_2=1.7\pm 0.3$ \cite{102},
$D_2=1.6\pm 0.1$ \cite{28},  $D_2=1.33\pm 0.02$ \cite{103},
$\eta=1.2\pm 0.15$, $\eta=1.3\pm 0.2$, $\eta=1.5\pm 0.3$ \cite{25},
$D_2=1.68$ \cite{26}, $D_2=1.30\pm 0.05$ \cite{27}, $D_2=1.28\pm 0.02$
\cite{110}) but in general show violation of an equality
$\eta=1$, which follows from the analysis of Ref.\,2. It is possible to
consider this fact as an evidence of hidden symmetry neglected in \cite{2}.
However, few critical remarks can be made in relation with the
cited researches.
\vspace{2mm}

(a) The analysis in \cite{23,25} is based on the relation between the
Fourier transform  of the density correlator (5) and the
diffusion coefficient
$$
S(\omega, q) \sim
\frac{D(\omega, q) q^2}{\omega^2+[D(\omega, q) q^2]^2} \,,
\eqno(15)
$$
which is valid only for real $D(\omega, q)$. The
general relation can be found in \cite{2} and is given
there by Eq.\,31:
$$
S(\omega, q) \sim \frac{1}{\omega} {\rm Im}
\frac{D(\omega, q) q^2}{-i \omega+D(\omega, q) q^2} \,.
\eqno(16)
$$
It is clear from the above
discussion that the low frequency diffusion coefficient is real
in the metallic state and pure imaginary in the localized
phase, while a complicated rearrangement of its analytical
structure  occurs
in  the vicinity of the transition point.
Thus, the most delicate feature of the Anderson transition is
completely ignored in \cite{23,25}.

In fact, only one real quantity $S(\omega, q)$ is measured in
numerical experiments, and one cannot extract two functions
${\rm Re} \,D(\omega, q)$ and ${\rm Im}\, D(\omega, q)$ from such
a measurement\,\footnote{\,In principle, one can use the
Kramers--Kronig relation and obtain a closed system of equations
for these quantities, but the contemporary algorithms are
far from fulfilling this program. }.
We see, that numerical experiments do not provide any direct
information on the diffusion coefficient. All conclusions
concerning $D(\omega, q)$ are based on strong assumptions
and cannot be considered as reliable.

\vspace{2mm}

(b) In our opinion, there
is some problem with the relation $\eta=d-D_2$.
According to Wegner \cite{100}, $D_2=2-\epsilon$ in the
$(2+\epsilon)$-dimensional case, suggesting $\eta=2\epsilon$.
However, Wegner's result was obtained for the 'minimal'
$\sigma$-model, where the spatial dispersion of $D(\omega, q)$
is neglected (see Sec.\,2 above); it looks as internal
inconsistency.

\vspace{2mm}

(c) The fractal dimension $D_2$ is determined by the behavior
of the participation ratio $P$ as a function of the system size $L$,
$P\sim L^{D_2}$. Theoretically, such behavior is expected only
for the critical disorder $W_c$, but in practice the power law
dependences $P\sim L^\alpha$ are observed for wide range of
disorder \cite{28} (see a detailed discussion in \cite{111}). In order the
measurement of $D_2$ was possible, the value $W_c$ is taken from other
experiments, which are not directly related with the properties of wave
functions. Analysis of the contemporary situation shows \cite{15} that
essential revision of the conventional results for $W_c$ is practically
enevitable. As a result, a value $D_2$ can change essentially.

\vspace{3mm}

4. Relation of the diffusion poles with the Anderson transition
was discussed recently in the  series of papers
by Janis and Kolorenc [28--31],
where the following conclusions were reached:

(i) Existence of the diffusion pole in the localized phase is
incompatible with the Ward identity;

(ii) The diffusion pole is absent in the localized phase;

(iii) The Ward identity is violated due to the averaging procedure.

\noindent
The initial point for these conclusions is the solution for the
two-particle Green function obtained for high dimensionalities
of space $d$, which (according to the authors) is asymptotically exact
in the parameter $d^{-1}$ \cite{16}. This solution does not posess the
diffusion pole in the localized phase and does not satisfy
the Ward identity. Instead to seek an origin of this surely
defective result, the authors of [28--31]
become
anxious to defend it. Using reasonable, but unjustified form
of $U_{{\bf kk'}} ({\bf q})$, they came to conclusion on the
principal impossibility to satisfy the Ward identity in the
localized phase \cite{17}. Its violation is argued by
possibility that the Hilbert space of the eigenfunctions
becomes incomplete due to the averaging procedure \cite{18}.

\vspace{2mm}

We have specially stressed in Sec.\,1, that in order to satisfy the Ward
identity one needs very delicate approximations: they should be
not only asymptotically exact in some parameter but should be made
'in the proper place'\,\footnote{\,There is nothing new in these
considerations. It is well known that the first correction to the
self energy has essentially more sense that the first correction
to the Green function; the one-loop approximation for the Gell-Mann - Low
function is more effective than the one-loop approximation to
the invariant charge, and so on. }. We see nothing unusual that
the results for high dimensions are not self-consistent, even if
they are really correct in the leading order in $d^{-1}$. Indeed, the
formal expansion of the right-hand side of Eq.\,8 in powers of $d^{-1}$ may
contain the well defined zero order term, while divergent integrals may occur
in the higher orders. Consequently, the result of \cite{16} for high dimensions
does not provide the reliable foundation for further conclusions.

\vspace{2mm}

The statement (i) on the principal impossibility of satisfying
the Ward identity in the localized phase (based on approximate
structure of $U_{{\bf kk'}} ({\bf q})$) contradicts the rigorous
analysis of the paper \cite{2}. The $1/\omega$ singularity
in  the right-hand side of Eq.\,8 is enevitably cancelled if the
spectral properties
of the quantum collision operator $\hat L$ are taken into account.
\vspace{2mm}

The statement (ii) on the absense of the diffusion pole in the
localized phase contradicts the existence of the Berezinskii-Gorkov
singularity $1/\omega$, which can be obtained not only
in the framework of the self-consistent theory \cite{1}  (which is
cited in [28--31])
but by direct analysis of the density
correlator \cite{3,4,5,2} and in the framework of the instanton
approach \cite{4,6}.
This singularity is closely related with the Anderson criterion
of localization \cite{19} in modification by Economou and Cohen
\cite{19a} (see \cite{4} and \cite{6}), which can be
rigorously proved for $1D$ systems \cite{20}. In fact, the existence
of the $1/\omega$ singularity can be easily seen for a zero-dimensional
case, which corresponds to the Anderson model on a single site. The
exact Green function
$$
G^R(E) =\frac{1}{E-V+i 0}
\eqno(17)
$$
after trivial averaging takes a form
$$
\langle G^R(E)\rangle = \int\limits_{-\infty}^{\infty}
\frac{P(V)\, dV}{E-V+i 0}
\eqno(18)
$$
($P(V)$ is a distribution of $V$), giving a density of states
$$
\nu(E)= -\frac{1}{\pi} {\rm Im}\,\langle G^R(E)\rangle =
P(E)\,.
\eqno(19)
$$
Then the quantity $\phi^{RA}$,
$$
\phi^{RA}= \langle G^R(E+\omega) G^A(E) \rangle =
\int\limits_{-\infty}^{\infty} \frac{1}{E-V+\omega+i 0}\,
\frac{1}{E-V-i 0}\, P(V)\, dV \,,
\eqno(20)
$$
has the $1/\omega$ singularity for $\omega\to 0$,
$$
\phi^{RA}= \frac{2\pi P(E)}{-i\omega}\,,
\eqno(21)
$$
in accordance with Eq.\,3 (the momentum dependence
is absent in the zero-dimensional case, and all momenta are supposed
to be equal to zero; so $q=0$ and $A(0)=1$). The results (17--21) are
also valid for a space of arbitrary dimensionality in the limit
of strong disorder. Already these simple results are not reproduced
by sophisticated expressions in [28--31].

It is well known, that the coherent potential approximation (CPA),
considering as formally exact in the leading order
in $d^{-1}$, has an essential qualitative defect \cite{32}: it does not
describe
the fluctuational tail of the density of states, which
is exponentially small and cannot be found in any finite order
of perturbation theory.
This tail can be explicitly calculated by the instanton method both
for $d<4$ \cite{6,7,33} and for higher dimensions
\cite{36,13}\,\footnote{\,Absence of instantons for $d>4$ \cite{37} should not
be considered as the absence of localized states \cite{38}. The theory is not
renormalizable for $d>4$ and should be considered on the lattice.  The lattice
instantons always exist. }.  According to \cite{4,6}, the Berezinskii-Gorkov
singularity has also nonperturbative origin, and
it is naturally not reproduced by the extention of CPA used in [28--31].

\vspace{2mm}

The statement (iii) on the possible violation of the Ward identity
due to the averaging procedure \cite{18} is based mainly on
philosophical considerations. In principle, such argumentation
may be correct in some situations, but it has nothing to do with
the present case: the Ward identity (8) was diagrammatically derived
in \cite{1} directly for the averaged quantities. Of course, one
can discuss the possible nonperturbative contributions, but
at the present case there is no ground for them: validity of the
Ward identity corresponds to the physical requirements, and its
violation is not expected. In
fact, the mystical fear of nonperturbative contributions is not grounded:
such contributions can be safely extracted from the diagrammatic expansions
\cite{13} if a correct procedure for summing of divergent series is chosen
\cite{21}.

\vspace{2mm}

One can see, that
all statements (i),(ii),(iii) are incorrect.
The authors of [28--31]
were wrecked on the 'hidden rocks',
which were discovered many years ago.

\vspace{4mm}

This work is partially supported by RFBR  (grant 03-02-17519).



\begin{thebibliography}{xx}

\bibitem{1} D. Vollhardt and P. W${\rm {\ddot o}}$lfle, Phys. Rev. B {\bf
22}, 4666 (1980); Phys. Rev. Lett. {\bf 48}, 699 (1982); in Modern
Problems in Condensed Matter Sciences, edited by V. M. Agranovich and A.
A. Maradudin, North-Holland, Amsterdam, 1992, Vol. 32.

\bibitem{2} I. M. Suslov, Zh. Eksp. Teor. Fiz. {\bf 108}, 1686 (1995)
[JETP {\bf 81}, 925 (1995)]; cond-mat/0111407.

\bibitem{3} V. L. Berezinskii and L. P. Gor'kov, Zh. Eksp. Teor. Fiz. {\bf
77}, 2498 (1979) [Sov. Phys. JETP {\bf 50}, 1209 (1979)].

\bibitem{4} M. V. Sadovskii, Zh. Eksp. Teor. Fiz. {\bf 83}, 1418 (1982)
[Sov. Phys. JETP {\bf 56}, 816 (1982)].

\bibitem{5} M. V. Sadovskii, Sov. Sci. Rev. A. Phys. {\bf 7}, 1 (1986).

\bibitem{8} K. B. Efetov, Zh. Eksp. Teor. Fiz. {\bf 88}, 1032 (1985) [Sov.
Phys. JETP {\bf 61}, 606 (1985)].

\bibitem{9} B. L. Al'tshuler, A. G. Aronov, D. E. Khmel'nitskii, and A. I.
Larkin in Quantum Theory of Solids, edited by I. M. Lifshitz, Mir
Publishers, Moscow, 1982.

\bibitem{10} F. Wegner, Z. Phys. B {\bf 35}, 207 (1979);
L. Sch$\ddot a$fer, F. Wegner, Z. Phys. B {\bf 38}, 113 (1980).
S. Hikami, Phys. Rev. B {\bf 24}, 2671 (1981).

K. B. Efetov, A. I. Larkin, D. E. Khmelnitskii, Zh. Eksp. Teor. Fiz.
{\bf 79}, 1120 (1980) [Sov. Phys. JETP {\bf 52}, 568 (1980)].

K. B. Efetov, Adv. Phys. {\bf 32}, 53 (1983).

\bibitem{11} V. E. Kravtsov, I. V. Lerner, V. I. Yudson,
Zh. Eksp. Teor. Fiz. {\bf 94}, 255 (1988)
 [Sov. Phys. JETP {\bf 67}, 1441 (1988)].

 F. Wegner, Z. Phys. B {\bf 78}, 33 (1990).

\bibitem{12} S. Ma, Modern Theory of Critical Phenomena, Reading, Mass.:
W.A.Benjamin, Advanced Book Program, 1976.

\bibitem{13} I. M. Suslov, Usp. Fiz. Nauk {\bf 168}, 503 (1998)
[Physics-Uspekhi {\bf 41}, 441 (1998)]; cond-mat/9912307.


\bibitem{34} D. Belitz, T. R. Kirkpatrick, Rev. Mod. Phys., {\bf 66}, 261
(1994).

\bibitem{35} N. G. Zhdanova, M. S. Kagan, E. G. Landsberg,
JETP {\bf 90}, 662 (2000).  This experiment is made on the nondegenerate
electron gas and there is no question on influence of interaction.

\bibitem{24} E. Abrahams, P. W. Anderson, D. C. Licciardello, and T. V.
Ramakrishman, Phys. Rev. Lett. {\bf 42}, 673 (1979).

\bibitem{14}  P. Markos, cond-mat/0609580.

\bibitem{15} I.~M.~Suslov, cond-mat/0610744.

\bibitem{22} E.~Abrahams, P.A.Lee, Phys.~Rev.~B {\bf 33}, 683
(1986).

\bibitem{23} J.~T.~Chalker, Physica~A {\bf 167}, 253
(1990).

\bibitem{25} T. Brandes, B. Huckestein, L. Schweitzer,
Ann. Phys. {\bf 5}, 633 (1996).

\bibitem{102}  C. M. Soukolis, E. N. Economou, Phys. Rev. Lett.
{\bf 52}, 565 (1984).

\bibitem{28} M.~Schreiber, J.~Phys.~C {\bf 18}, 2490 (1985);
Physica~A {\bf 167}, 188 (1990).

\bibitem{103} S. N. Evangelou, Physica A {\bf 167}, 199 (1990).

\bibitem{26} M.~Schreiber, H.~Grussbach, Phys.~Rev.~Lett. {\bf 67}, 607
(1991).

\bibitem{27} A.~Mildenberger, F.~Evers, A.~D.~Mirlin, Phys.~Rev.~B {\bf 66},
033109 (2002).

\bibitem{110} J. Brndiar, P. Markos, cond-mat/0606056.


\bibitem{100} F. Wegner, Z. Phys. B {\bf 36}, 209 (1980).

\bibitem{111} I. M. Suslov, Zh. Eksp. Teor. Fiz. {\bf 129}, 1064 (2006)
[JETP {\bf 102}, 938 (2005)]; cond-mat/0512708.

\bibitem{16}  V. Janis, J. Kolorenc,  Phys. Rev. B {\bf 71},
033103 (2005), cond-mat/0402471.

\bibitem{17}  V. Janis, J. Kolorenc, Phys. Stat. Sol. B {\bf 241},
2032 (2004), cond-mat/0403618.


\bibitem{18}  V. Janis, J. Kolorenc,  cond-mat/0407618.

\bibitem{18a}  V. Janis, J. Kolorenc, Phys. Rev. B {\bf 71},
245106 (2005), cond-mat/0501586.

\bibitem{6} J. L. Cardy, J. Phys. C {\bf 11},  L321 (1978).

\bibitem{19} P. W. Anderson, Phys. Rev. {\bf 109}, 1492 (1958).

\bibitem{19a}  E. N. Economou, M. H. Cohen, Phys. Rev. B {\bf 5}, 2931 (1972).


\bibitem{20} K. Ishii, Progr. Theor. Phys. Suppl. {\bf 53}, 77 (1973).

\bibitem{32} D. J. Thouless, Phys. Rep. {\bf 13}, 92 (1974).

\bibitem{7} M. V. Sadovskii, Fiz. Tverd. Tela (Leningrad) {\bf 21},  743
(1979).

\bibitem{33} E. Brezin, G. Parisi, J. Phys. C {\bf 13}, L307 (1980).

\bibitem{36} I. M. Suslov, Zh. Eksp. Teor. Fiz. {\bf 102}, 1951 (1992)
[Sov. Phys. JETP {\bf 75}, 1049 (1992)];
Zh. Eksp. Teor. Fiz. {\bf 105}, 560 (1994) [JETP {\bf 79}, 307 (1994)].

\bibitem{37} V. G. Makhankov, Phys. Lett. {\bf 61A}, 431 (1977).

\bibitem{38} D. J. Thouless, J. Phys. C {\bf 9}, L603 (1976).

\bibitem{21} I. M. Suslov, Zh. Eksp. Teor. Fiz. {\bf 127}, 1350 (2005)
[JETP {\bf 100}, 1188 (2005)]; cond-mat/0510142.




\end{thebibliography}
\end{document}